\documentclass[doublecol]{epl2}
\usepackage{amsmath}
\usepackage{amssymb}

\title{Disorder and magnetic transport in tilted Weyl semimetals}
\author{Yi-Xiang Wang$^{1,2}$ }
\institute{
$^1$School of Science, Jiangnan University, Wuxi 214122, China.\\
$^2$Department of Physics and Astronomy, University of Pittsburgh, Pittsburgh, Pennsylvania 15260, USA
}

\abstract{
We investigate the effect of disorder on the Landau levels (LLs) in tilted three-dimensional Weyl semimetals (WSMs) when a magnetic field is present.  Based on the minimum lattice model and by using the exact diagonalization and Kubo's formula, we numerically calculate the Hall conductivity and the density of states (DOS), from which several striking signatures are found to distinguish type-I WSMs from type-II WSMs: the first is the response of the Hall conductivity to the Fermi energy around the band center in clean limit, the second is the performance of the Hall conductivity to disorder, where in type-I WSMs, the robustness of the low-energy LLs is broken successively from the higher LLs to the lower ones and can be understood with the \textit{sink down} picture, and the third is the behavior of the DOS at zero energy to disorder.  The implications of our results are discussed. 
}

\begin{document}
\maketitle

\section{Introduction}

In the past few years, the study of topological phases of matter has been extended beyond the gapped states and now also includes various gapless nodal systems.  Three-dimensional (3D) Weyl semimetals (WSMs) are the prime examples of such gapless system \cite{N.P.Armitage}, whose characteristic properties lead to the protected Fermi arcs surface states and novel chiral anomaly in response to the external electromagnetic fields \cite{X.Wan}.  As the strict symmetries of free space do not necessarily hold in a crystal, new types of fermions may emerge in condensed matter background.  In high-energy physics, the Weyl cone tilting is usually forbidden by the Lorentz symmetry, but it can generically appear in a linearized low-energy theory around an isolated twofold band-crossing point in a crystal \cite{A.A.Soluyanov}.  If the tilting is weak that the Fermi surface remains pointlike, the system is classified as a type-I WSM (WSM1).  While if the tilting becomes strong, the Fermi surface may instead consist of the electron and hole pockets.  In this case, the system is classified as a type-II WSM (WSM2) \cite{Y.Xu}.  It is likely that some materials can undergo the transitions from WSM1 to WSM2 with doping or under pressure.  The experimental evidences for the WSM2 state have been reported recently in MoTe$_2$ \cite{L.Huang, K.Deng, J.Jiang, A.Tamai}, WTe$_2$~\cite{C.Wang}, and the alloy Mo$_x$W$_{1-x}$Te$_2$~\cite{I.Belopolski2016a, I.Belopolski2016b}.  Besides the Fermi arcs, WSM2 also owns an additional class of the so-called track states \cite{T.M.Mccormick}, which are nontopological but degenerate with the Fermi arcs and have been demonstrated in experiment \cite{L.Huang}. 

When the WSM system is disordered, a common viewpoint dominates that disorder plays an indispensable role in determining the properties of the system and may even drive the phase transition into the diffusive metal state \cite{H.Shapourian, M.J.Park, Y.Wu, C.Z.Chen}.  The previous studies show that when there is no tilting, the vanishing density of states (DOS) at Weyl nodes can persist up to a finite value of disorder strength, beyond which a metallic state is driven by disorder \cite{J.H.Pixley, K.Kobayashi, S.Bera}.  The critical properties of the semimetal-metal transition have also been analyzed \cite{P.Goswami, S.V.Syzranov, T.Louvet, B.Sbierski}.  In Ref.~\cite{J.H.Wilson}, it was demonstrated that the Weyl Fermi arcs are not topologically protected when disorder sets in, but the surface chiral velocity is robust and survives in the presence of strong disorder. 

One significant feature of WSMs is that the system can reach the quantum limit where the Landau levels (LLs) are formed even in a weak magnetic field \cite{K.Y.Yang, L.P.He, S.Jeon, R.Y.Chen, Y.X.Wang2017b}.  Consequently, WSMs provide an ideal platform to investigate the 3D Hall physics.  The studies of magneto-optical properties revealed that the absorption peaks of optical conductivity can give evident signals of WSM2 \cite{Z.M.Yu, M.Udagawa, S.Tchoumakov}.  To our knowledge, the effects of disorder on the LLs and magnetic transport properties in 3D tilted WSMs are less studied.  In this paper, based on the minimum lattice model, we try to investigate the interplay between disorder and LLs in tilted WSMs.  

As the quantum Hall effect (QHE) is a hallmark in two-dimensional (2D) electron system, 3D system normally does not exhibit the QHE.  This is due to the dispersive bands along the direction of the magnetic field, which can smear the energy gap between the LLs.  In recent experiments, due to the delicate quantum confinements, the QHE in 3D Dirac semimetal Cd$_3$As$_2$ thin films has been successfully observed \cite{M.Uchiba, S.Nishihaya, T.Schumann, C.Zheng, B.C.Lin}, but its origin is in debate and no consensus has been reached yet.  Several mechanisms were proposed for the QHE that it may be attributed to the quantum confinement induced bulk subbands \cite{M.Uchiba, S.Nishihaya}, the Weyl orbits that connect the opposite surfaces via bulk Weyl nodes\cite{T.Schumann, C.Zheng}, as well as the  topological-insulator-type surface states \cite{B.C.Lin}.  Here we only consider the underlying 3D LL physics as the stacked 2D quantum Hall system. 

In this paper, by using the exact diagonalization and Kubo's formula,  we calculate the Hall conductivity and the DOS in tilted WSMs.  Several striking signatures are found to distinguish WSM1 and WSM2.  The first is the response of the Hall conductivity to the Fermi energy around the band center in clean limit: the slope of the Hall conductivity vs the Fermi energy is vanishingly small in WSM1, but will jump to be much large in WSM2.  The second is that in WSM1, the Hall conductivity at low energy shows certain robustness to weak disorder and the robustness will be broken successively from the higher LLs to the lower ones when disorder increases, which can be understood by the \textit{sink down} picture.  While in WSM2, the Hall conductivity at low energy is more fragile to disorder and the robustness is absent.  The third is the behavior of the DOS at zero energy, where in WSM1, it remains to be quite small until disorder increases to the critical value, beyond which the system is driven into the diffusive metallic state, but in WSM2, it decreases continuously with disorder.  These observations demonstrate that the low-energy LLs in the energy window are robust to disorder, but the high-energy LLs outside the energy window are not.  Our work may help deepen the understanding of the interplay between disorder and magnetic fields in 3D tilted WSMs.

\section{Model}

We start from the minimum model Hamiltonian that supports a pair of Weyl fermions \cite{M.J.Park, H.Shapourian, Y.X.Wang2017a, M.Udagawa,Y.Wu},
\begin{align}
H_0=&2t(\text{sin}k_y\sigma_y+\text{sin}k_z\sigma_z)+(m_1-2t\text{cos}k_x)\sigma_x
\nonumber\\
&+m_0(2-\text{cos}k_y-\text{cos}k_z)\sigma_x
+2t_z\text{sin}k_z\sigma_0. 
\end{align}  
Here $\mathbf\sigma$ are the Pauli matrices acting on the spin space, $t$ is the hopping integral and the wave vector $k_i$ is measured by $1/a_0$ with $a_0$ being the lattice spacing.  $m_1$ controls the positions of Weyl nodes and $m_0$ denotes the Wilson mass term as to open a finite energy window to avoid the band overlapping.  The last term specifies the Weyl cone tilting in $z-$direction.  When the tilting is absent, $t_z=0$, $H_0$ preserves the inversion symmetry but breaks the time-reversal-symmetry (TRS) of the system.  For $|m_1|<2t$, the two Weyl nodes are given as $\boldsymbol K_\eta=\Big(\eta\text{arccos}(m_1/2t),0,0\Big)$, with $\eta=\pm$.  The introduction of the tilting term in $z-$direction that is perpendicular to the distance between the Weyl nodes breaks the inversion symmetry.  More importantly, it can easily tune the Weyl cone tilting to feature the electron and hole pockets.  Around Weyl node $\boldsymbol K_\eta$, the Hamiltonian $H_0$ is expanded to yield a low-energy continuous description, 
\begin{align}
H_{0\eta}(\boldsymbol k)=\hbar v(\eta k_x\sigma_x+k_y\sigma_y+k_z\sigma_z)+\hbar v_zk_z, 
\end{align}
with the velocities $v=2t/\hbar$ and $v_z=2t_z/\hbar$.  In the following, $t$ is set as the unit of energy, and the mass parameters are taken as $m_1=0$ and $m_0=2$.

We concentrate on a magnetic field along the tilting direction of the Weyl cones.  Such a magnetic field can lead to the dispersive LLs only along the $z-$direction and the flat LLs in the x-y plane. If the magnetic field acts in the $x-y$ plane, the chiral LLs are missing \cite{M.Udagawa}.  The magnetic field $\boldsymbol B=(0,0,B)$ can be included by the Peierls substitution, $\boldsymbol p\rightarrow\boldsymbol p-e\boldsymbol A$, with the vector potential in the Laudau gauge of ${\boldsymbol A(\boldsymbol r)}=(-yB,0,0)$.  Then the energy of the $n-$LL is obtained as
\begin{align}
&\varepsilon_{n\ge 1}(k_z)=\text{sgn}(n)
\sqrt{\hbar^2v^2k_z^2+\frac{2\hbar^2v^2|n|}{l_B^2}}+\hbar v_zk_z,
\\
&\varepsilon_{n=0,\eta}(k_z)=\hbar(\eta v+v_z)k_z, 
\end{align}
with the magnetic length $l_B=\sqrt{\hbar/eB}$.  
 
\begin{figure}
	\includegraphics[width=8.8cm]{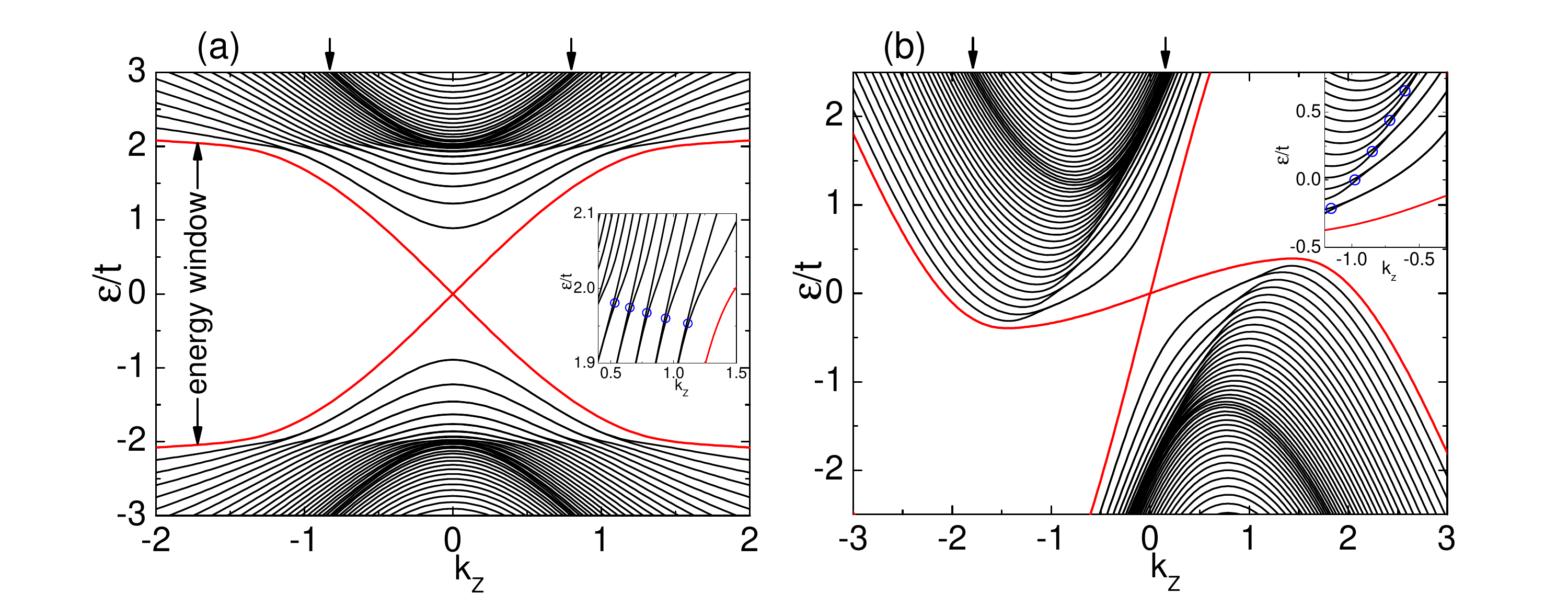}
	\caption{(Color online) The LL energy spectra of WSMs are obtained from the discretized lattice model with magnetic flux $p=60$, (a) $t_z=0$ and (b) $t_z=1.5$.  The chiral $n=0$ LLs are labeled with red color and {\color{red} the energy window is also specified}.  The arrows in the above point out the Van-Hove singularities where the bands are very dense and the DOS are divergent.  The insets show that the LL splittings exhibit the same characteristics as they cross the energy window even if the Weyl cones are overtilted. }
	\label{LLs}
\end{figure}
 
Eq.~(3) shows that $n\geq1$ LLs have the same energies in both Weyl nodes, so they are twofold degenerate.  The degeneracy is broken when the LLs cross the energy window, as shown by the circles in the insets of fig.~\ref{LLs} of the LL splittings.  We can also see that $n=0$ chiral LLs are clearly distinguishable in two Weyl cones and are independent of the magnetic field.  The Weyl cone tilting can drastically change the properties of the chiral $n=0$ LLs \cite{M.Udagawa}.  If $t_z<t$ and the system lies in WSM1, the two chiral $n=0$ modes in different Weyl cones own opposite velocities and are counter-propagating, as shown in fig.~\ref{LLs}(a).  If $t_z$ increases to $t_z>t$, the Weyl cones are overtilted and the system lies in WSM2, the two chiral modes acquire the velocities in the same direction, as in fig.~\ref{LLs}(b).

The tight-binding model can be discretized on a cubic lattice with the periodic boundary conditions in all directions.  In the lattice model, the magnetic field is written as $B=\phi/a_0^2\cdot h/e$ with the help of the dimensionless quantity $\phi$, which gives the magnetic flux penetrating a unit cell in $x-y$ plane and is in unit of the elementary flux quantum $h/e$ \cite{Y.X.Wang2014}.  For convenience, we take $\phi=1/p$ in which $p$ is an integer so that the magnetic flux will be commensurate with the lattice structure.  We take the model as a cubic lattice of $L_x=L_y=L_z=p$.  fig.~\ref{LLs} shows the LL energy spectra of WSMs obtained from the discretized lattice model with the magnetic flux $p=60$. 

In this paper, for WSM1, we term the LLs in the energy window as the ``low-energy LLs", as specified in fig.~\ref{LLs}(a), and those outside the energy window as the ``high-energy LLs".  When the Weyl cones are overtilted, the topological properties of the LLs will keep unchanged, including the characteristic Chern number and the splittings when the LLs cross the energy window (see the insets in fig.~\ref{LLs}).  So we still use the low-energy and high-energy LLs to distinguish the different LLs in WSM2.  It has been checked that the low-energy LLs in fig.~\ref{LLs} are consistent with the continuous model, as given in eqs. (3) and (4).

The disorder effect is introduced by adding the random on-site potential $H_{\text{dis}}$ \cite{H.Shapourian,M.J.Park,Y.Wu,C.Z.Chen} to the discretized lattice model,
\begin{align}
H_{\text{dis}}=\sum_{x,y,k_z,s}
\epsilon_{x,y,k_z,s}c^\dagger_{x,y,k_z,s}
c_{x,y,k_z,s}, 
\end{align}
where $x/y$ and $s$ are the coordinate and spin index, respectively.  $\epsilon$ is a random number uniformly distributed in the range $[-W/2,W/2]$, with $W$ being the disorder strength.  As the disorder configurations do not preserve the TRS, both the charge and magnetic disorder are included.  To make the exact-diagonalization calculations tractable, we consider the quasi-disorder case, \textit{i.e.}, the disordered system keeps the translational symmetry along $z-$direction.  Such a quasi-disorder case can represent the completely random disorder to some extent and was previously used to investigate the one-parameter scaling behavior at the Dirac point in graphene \cite{J.H.Bardarson} and the disorder-induced phase transitions from WSM1 to WSM2 \cite{M.J.Park}.  In next two sections, we will calculate the Hall conductivity and the DOS.  As disorder sets in the system, we need to do certain configuration average with the number $N_c=10^3\sim10^4$ to ensure the convergence of the numerical results.  More details can be found in the Appendix.

\section{Hall conductivity}

\begin{figure}
	\includegraphics[width=8.8cm]{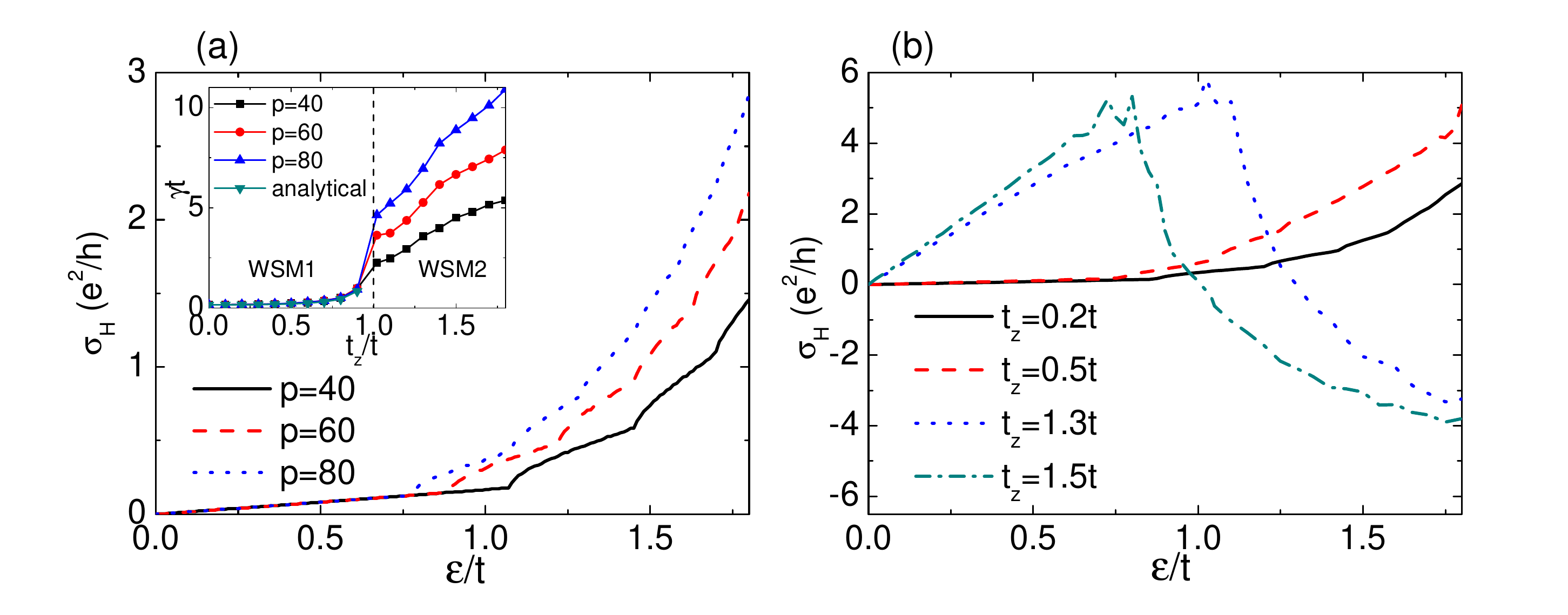}
	\caption{(Color online) The Hall conductivity $\sigma_H$ vs the Fermi energy $\varepsilon$ around the band center for WSMs under the magnetic field.  (a) is for different magnetic flux $p$ and no tilting.  (b) is for different tilting factors $t_z$ with $p=60$.  The inset in (a) shows the slope $\gamma$ vs $t_z$ around zero energy, where in WSM1, the numerical results are in good agreement with the analytical results.}
	\label{sigmaH}
\end{figure}

As we consider the 3D LL physics as the stacked 2D quantum Hall system, the Hamiltonian $H(\boldsymbol k)$ is regarded as a superposition of the 2D slices $H(k_x,k_y)$ in $z$-direction, with the mass term $m(k_z)$.  Then $\sigma_H$ is expressed as a function of the Fermi energy $\varepsilon$ \cite{A.A.Burkov},
\begin{align}
\sigma_H(\varepsilon)=\int_{-\pi}^\pi\frac{dk_z}{2\pi} \sigma_H^{2D}(\varepsilon,k_z).
\end{align}
Here the 2D Hall conductivity $\sigma_H^{2D}$ at zero temperature is calculated with the Kubo's formula \cite{D.J.Thouless, Y.X.Wang2014}. 

First of all, we investigate the Hall conductivity $\sigma_H$ in the disorder-free WSM system.  Because the LLs are dispersive in 3D WSMs and almost indiscernible near the band edge, we focus on $\sigma_H$ around the band center.  The numerically calculated Hall conductivity on the lattice model is shown in fig.~\ref{sigmaH} with the given parameters.  

Actually, when $t_z<t$ and the Fermi energy $(\varepsilon>0)$ lies in the low-energy region, $\sigma_H$ is given analytically as
\begin{align}
\sigma_H=&\frac{e^2}{h}\frac{v}{\pi\hbar(v^2-v_z^2)}
\Big[\varepsilon+\sum_{n\geq1}2\theta(\varepsilon>\varepsilon_n^v)\sqrt{\varepsilon^2-\varepsilon_n^{v,2}} \Big], 
\label{anasigmaH}
\end{align}
here $\varepsilon_n^v=\hbar/l_B\cdot\sqrt{2n(v^2-v_z^2)}$ is the vertex energy of the dispersive $n-$LL and $\theta(x)$ is the step function.  In eq.~(\ref{anasigmaH}), the first term is due to the contributions from $n=0$ LLs, which for specific $k_z$ carry Chern number 1, and the second term comes from $n\geq1$ LLs, which for specific $k_z$ carry Chern number 2 due to the degeneracy of the LLs in the two Weyl nodes.  As the particle-hole symmetry is preserved, similar results for $\sigma_H$ with negative Fermi energy can also be obtained. 

Eq.~(\ref{anasigmaH}) tells us that $\sigma_H$ increases with $\varepsilon$ as the LLs of larger momentum region are occupied by the electronic states.  Each time the Fermi energy crosses the vertex energy $\varepsilon_n^v$ of the higher $n-$LL, the $n-$LL begins to contribute to $\sigma_H$.  This makes $\sigma_H$ seem like folding lines, as demonstrated by the numerical results in fig.~\ref{sigmaH}(a) with $t_z=0$.  Note that the larger $p$ corresponds to the smaller magnetic flux, so more folding lines in $\sigma_H$ appear.  Comparing the different lines in fig.~\ref{sigmaH}(a), we can see that when $p$ increases, $\sigma_H$ also increases, as more LL states are occupied for certain Fermi energy $\varepsilon$.  This observation agrees with previous works, as in fig. 1(b) of Ref.~\cite{H.W.Wang} and in fig. 10 of Ref.~\cite{V.Konye}.  It is worth emphasizing that the study of $\sigma_H$ here lies in the quantum oscillation regime, or regime II of Ref.~\cite{H.W.Wang}.  In the classical picture, when the magnetic field is strong enough and the temperature is very low, the Hall conductivity is given as $\sigma_H=-\frac{ne}{B}$, with $n$ denoting the electron density.  This result is a consequence of the completely disorder-free limit that we consider here, in which the scattering time is infinite.  For a finite scattering time, one retrieves the usual behavior $\sigma_H\propto B$ in the small-$B$ field limit. 

When the Weyl cones are tilted as $t_z>0$, if the tilting is weak, $t_z<t$, the multiple folding lines still exist in $\sigma_H$, as shown in fig.~\ref{sigmaH}(b).  If the Weyl cones are overtilted, $t_z>t$, the system enters WSM2 and $\sigma_H$ includes the contributions from the low-energy LLs as well as the high-energy ones.  As the Fermi energy increases, the contributions to $\sigma_H$ from the lower Weyl cone are decreasing, while the contributions from the upper Weyl cone are increasing, leading to an approximate linear relationship of $\sigma_H$ with $\varepsilon$, as shown in fig.~\ref{sigmaH}(b).  When $\varepsilon$ increases to the lowest Van-Hove singularity of a specific LL, $\sigma_H$ reaches its maximum and then decreases rapidly to the negative value.  The positions of the Van-Hove singularities are shown by the arrows in fig.~\ref{LLs}, where the LLs are very dense and the DOS diverge.  In the extreme case that the Weyl cones are heavily overtilted, $t_z>2$, the Van-Hove singularities will move to be around zero energy, leading to the negative Hall conductivity (not shown here).

According to the above analysis, around the band center, the linear relationship can be fit between $\sigma_H$ and $\varepsilon$ as
\begin{align}
\sigma_H(t_z,\varepsilon)=\frac{e^2}{h}\gamma(t_z)\varepsilon.  
\end{align}
The slope $\gamma$ vs the tilting factor $t_z$ for different magnetic flux $p$ is plotted in the inset of fig.~\ref{sigmaH}(a).  It shows that in WSM1, the slope $\gamma$ is vanishingly small, while in WSM2, $\gamma$ suddenly jumps to a much larger value.  This is indeed expected as one has a large DOS at the Fermi level of WSM2, in contrast to the vanishing DOS of WSM1.  We can also see that in WSM1, the slope $\gamma$ is independent of the magnetic flux $p$, because the energies of $n=0$ LLs are independent of $p$, as in eq. (4).  The numerical results on the lattice model are in good agreement with the analytical results in eq.~(\ref{anasigmaH}), from which $\gamma=v/[\pi\hbar(v^2-v_z^2)]$.  In WSM2, $\gamma$ increases with $p$, as more LLs are occupied below the Fermi energy.  The different responses of $\sigma_H$ to the Fermi energy around the band center can be used as a signature to distinguish WSM1 from WSM2.  To observe the characteristic variance of Hall conductivity in experiment, one needs a clean WSM sample where the Fermi energy can be modulated by doping the system. 

\begin{figure}
	\includegraphics[width=8.8cm]{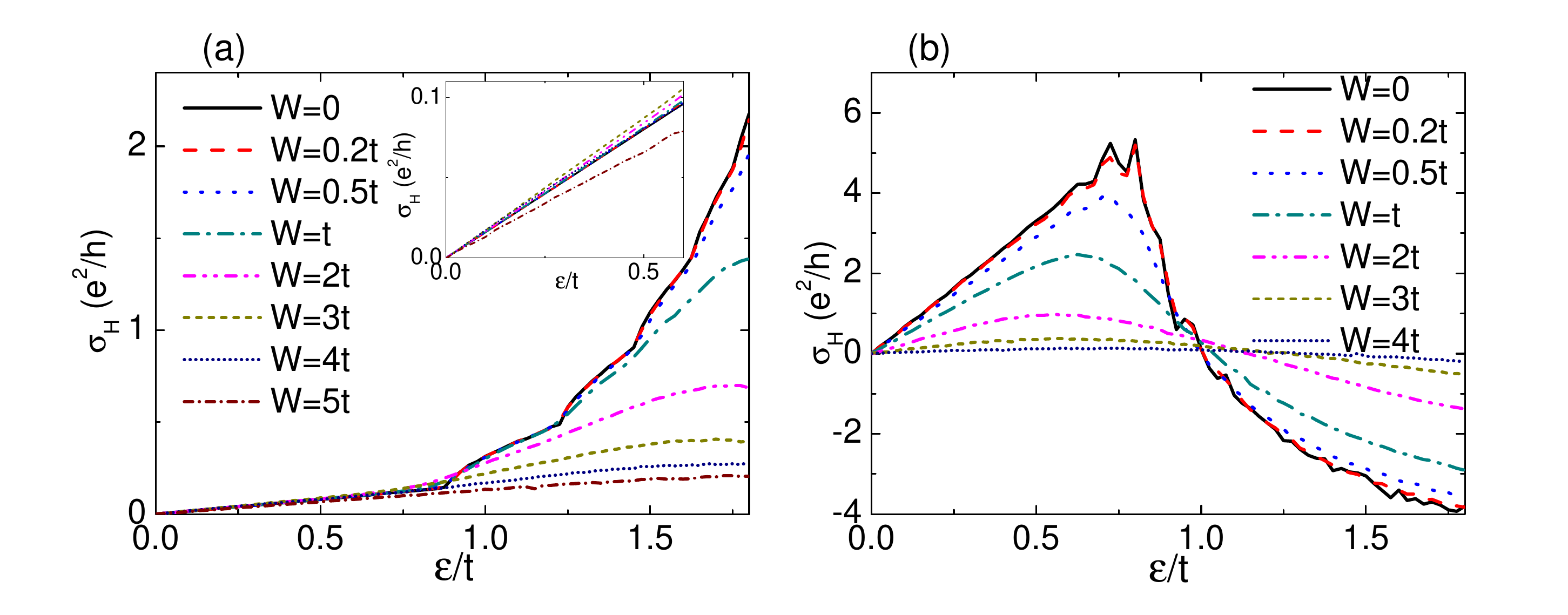}
	\caption{(Color online) The Hall conductivity $\sigma_H$ around the band center for different disorder strength $W$.  The tilting parameter is $t_z=0$ in (a) and $t_z=1.5$ in (b).  We set the magnetic flux $p=60$.  The inset in (a) shows an expanded view of $\sigma_H$ around zero energy.  }
	\label{sigmaH_dis}
\end{figure}

Next we turn to the effect of random disorder on the Hall conductivity in WSMs.  The results of $\sigma_H$ for different disorder strength $W$ are plotted in fig.~\ref{sigmaH_dis}.  We notice that $\sigma_H$ always vanishes at zero energy for all $W$, as the particle-hole symmetry is still preserved in a disordered system.  In fig.~\ref{sigmaH_dis}(a) with $t_z=0$, it shows when $W$ increases, $\sigma_H$ is suppressed successively from the higher LLs to the lower ones.  At $W=0.5t$, the contributions to $\sigma_H$ from $n=0,1,2,3$ LLs are well retained while the contributions from $n\ge4$ LLs are suppressed.  Further increasing the disorder strength and at $W=2t$, the contributions from $n\neq0$ LLs are suppressed.  Finally, at strong disorder $W=5t$, $\sigma_H$ due to $n=0$ LLs is suppressed, with the expanded view of $\sigma_H$ around zero energy being shown in the inset of fig.~\ref{sigmaH_dis}(a).  These numerical results suggest that in the low-energy region, the higher LLs are more susceptible to disorder while the lower LLs exhibit certain robustness.  The effects of disorder on $\sigma_H$ are also valid for weakly tilted WSM1.

These results can be explained as follows.  We label the wavefunction of one specific low-energy LL state as $\psi_{\varepsilon_1,k_{z1}}$ and the wavefunction of another specific state in the high-energy Van-Hove singularity regime as $\psi_{\varepsilon_2,k_{z2}}$.  As the two states own opposite Chern numbers, they will be annihilated if they meet, only when two conditions are satisfied: the momenta between the two states are equal $k_{z2}=k_{z1}$ and the energy difference $\Delta=\varepsilon_2-\varepsilon_1$ is compensated by the disorder-induced scatterings.  The compensation of the larger energy difference requires the stronger disorder strength $W$.  So the contributions to $\sigma_H$ are suppressed from the higher LLs at first, as the energy difference is small, and then with the increasing of disorder, gradually to the lower LLs and finally to $n=0$ LLs.  Thus for specific $k_z$, it seems like that the LLs carrying negative-Chern number coming from the higher energy region \textit{sink down} and are moving continuously towards the band center with increasing disorder strength.  When they cross the Fermi energy at the critical disorder strength $W_c$, they annihilate with the LLs carrying positive Chern number and the corresponding Hall conductivity is suppressed.　 These arguments remind us of the \textit{float up} picture in 2D quantum Hall system \cite{Th.Koschny, D.N.Sheng2001, D.N.Sheng2006}, which explains the collapse of the Hall plateau through the annihilation of positive-Chern number LLs around the band center and the \textit{float up} negative-Chern number LLs from the Van-Hove singularities due to the disorder-induced scatterings.  Note here we consider the positive Fermi energy, compared with the negative Fermi energy in their works \cite{Th.Koschny, D.N.Sheng2001, D.N.Sheng2006}.

\begin{figure}
	\includegraphics[width=8.8cm]{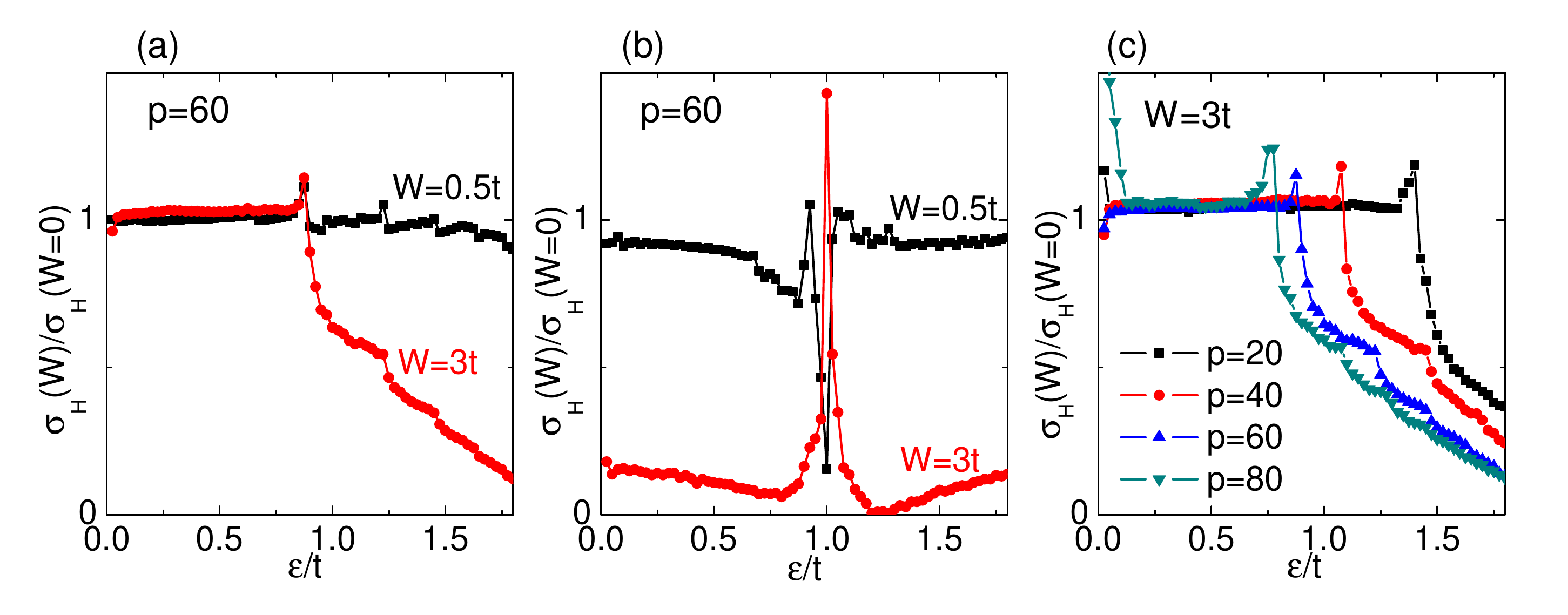}
	\caption{(Color online) The normalized Hall conductivity by the zero disorder limit $\frac{\sigma_H(W)}{\sigma_H(W=0)}$.  (a) and (b) are for different disorder $W$ in WSM1 with $t_z=0$ and WSM2 with $t_z=1.5$, respectively.  (c) is for different magnetic flux $p$ in WSM1 with $t_z=0$. }
	\label{norm_sigmaH}
\end{figure}

For WSM2, in fig.~\ref{sigmaH_dis}(b) with $t_z=1.5t$, it shows that $\sigma_H$ decreases gradually with disorder.  When the disorder strength reaches $W\sim4t$, $\sigma_H$ tends to vanish completely.  These numerical results suggest that $\sigma_H$ in WSM2 are more fragile to disorder.  We can attribute this to the fact that the overtilted Weyl nodes are concealed in the high-energy LLs.  As the high-energy LLs are not robust and easily broken by the disorder-induced scatterings (demonstrated by the DOS in the next section), it leads to the evident suppression of $\sigma_H$ in WSM2.  Thus the different responses of $\sigma_H$ to disorder can be used as the second signature to distinguish WSM1 from WSM2.

\begin{figure}
	\includegraphics[width=8.8cm]{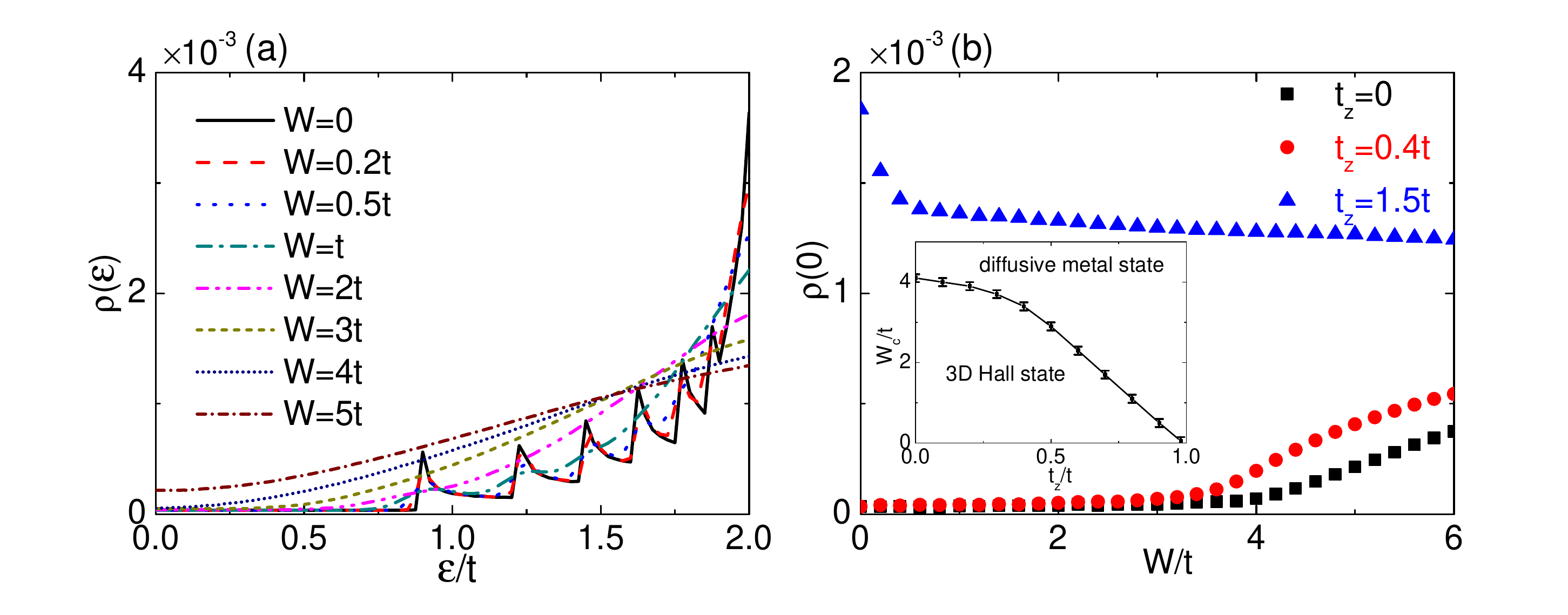}
	\caption{(Color online) (a) The DOS $\rho(\varepsilon)$ with $t_z=0$ around the band center for different disorder strength $W$.  (b) The DOS $\rho(0)$ vs disorder strength $W$ for different $t_z$.  The inset in (b) gives the critical disorder strength $W_c$ vs the weak $t_z$, which separates the 3D Hall state from the diffusive metal state.  We set the magnetic flux $p=60$. }
	\label{DOS}
\end{figure}

To further clarify how the difference between WSM1 and WSM2 depends on energy, we calculate the normalized Hall conductivity by the zero disorder limit $\frac{\sigma_H(W)}{\sigma_H(W=0)}$.  In figs.~\ref{norm_sigmaH}(a) and (b), when $\varepsilon$ is small, we compare the normalized $\sigma_H$ for different disorder $W$ in WSM1 and WSM2, respectively.  From these figures, the qualitatively different behaviors can be seen that the normalized $\sigma_H$ exhibits the non-monotonous behavior with disorder for WSM1, but decreases monotonously for WSM2.  Note that in WSM2, $\sigma_H$ is vanishing at $\varepsilon=t$ even if disorder sets in the system, so there exist large fluctuations of the normalized $\sigma_H$ around $\varepsilon=t$.

We also investigate the influence of the magnetic flux on the Hall conductivity in disordered WSM1.  In fig.~\ref{norm_sigmaH}(c), we plot the normalized $\sigma_H$ for different magnetic flux $p$, with a fixed disorder $W=3t$.  It shows that with the decreasing of $p$, the normalized $\sigma_H\simeq1$ can persist up to a larger Fermi energy, as it is closely related to $n=0$ LLs.  When the Fermi energy crosses the vertex of $n=1$ LL, the robustness of the normalized $\sigma_H$ will be broken by disorder.  This is because the smaller $p$ means the larger energy of $n=1$ LL, as in eq.~(3).  Then with the help of the above \textit{sink down} picture, the normalized $\sigma_H$ that is contributed by $n=0$ LLs for smaller $p$ can keep unaffected to a larger Fermi energy.

\section{Density of states}

To further investigate the effect of disorder on LLs, we calculate the normalized DOS in tilted WSMs,
\begin{align}
\rho(\varepsilon)=\frac{1}{N}\sum_{i=1}^N\delta(\varepsilon-\varepsilon_i),
\end{align}
with $N=2L_xL_yL_z/(a_0^3)$ being the total number of eigenvalues in the system.  When $t_z<t$ and $\varepsilon>0$ lies in the low-energy region, the DOS is obtained analytically as \cite{Y.X.Wang2017b},
\begin{align}
\rho(\varepsilon)
=\frac{1}{2\pi l_B^2}\frac{v}{\hbar(v^2-v_z^2)}\Big[1+\sum_{n\geq1} 2\theta(\varepsilon>\varepsilon_n^v)
\frac{\varepsilon}
{\sqrt{\varepsilon^2-\varepsilon_n^{v,2}}}\Big], 
\label{anaDOS}
\end{align}
here $\varepsilon_n^v$ is the same as in eq.~(\ref{anasigmaH}).  The first term in eq. (\ref{anaDOS}) comes from the chiral $n=0$ LLs and the second term is from $n\ge1$ LLs.   

The numerical results for the DOS with $t_z=0$ are plotted in fig.~\ref{DOS}.  In fig.~\ref{DOS}(a) when disorder is absent, the DOS exhibits a sawtooth shape with the square-root singularities at the vertex energy  $\varepsilon=\varepsilon_n^v$ from the one-dimensional dispersion of each LL.  With increasing disorder, the width of these peaks is broaden and their height is reduced.  If $W$ is beyond the critical value $W_c$, the corresponding peak in the DOS will disappear.  This also happens successively from the higher LLs at first and then to the lower ones and is consistent with the Hall conductivity analysis in the previous section.

We focus on the DOS at zero energy $\rho(0)$.  In WSM1 and with no disorder, $\rho(0)$ is due to the chiral $n=0$ LLs, which is a constant and rather small.  When disorder increases, $\rho(0)$ remains to be unaffected, as shown in fig.~\ref{DOS}(b) of $t_z=0$.  Further increasing disorder to beyond the critical value $W_c\sim4.2\pm0.2$, $\rho(0)$ will become finite and the system is driven into the diffusive metal state.  The observation also holds for $t_z=0.4$, with the critical disorder $W_c\sim3.6\pm0.2$.  So $\rho(0)$ can serve as an order parameter to characterize the continuous phase transition from the 3D Hall state to the diffusive metal state.  This is similar to the previous studies about the disordered WSMs without a magnetic field \cite{J.H.Pixley,S.Bera}, which revealed that for $t_z=0$, the critical disorder strength for $\rho(0)$ from vanishing to nonvanishing is around $W_c\sim2.55$ \cite{J.H.Pixley} and $W_c\sim3.3$ \cite{S.Bera}.  The critical values are much larger in the present study under a magnetic field, meaning that the magnetic field can make Weyl nodes become more stable to disorder.  In the inset of fig. \ref{DOS}(b), the critical $W_c$ vs $t_z<1$ is plotted, in which the uncertainty in $W_c$ arises in determining the exact position of the phase transition.  It shows that $W_c$ decreases with $t_z$ and tends to zero when $t_z\rightarrow1$.  This can be attributed to the competition between the Weyl cone tilting and the disorder that more electronic states from the high-energy regime can be scattered into the energy window when the Weyl cone tilting increases.  Note that the critical $W_c$ should also depend on the magnetic field and it may be scaled with the energy separation between the LLs, which will be left for the future work.
 
For comparison, in fig.~\ref{DOS}(b) for WSM2 with $t_z=1.5$, we can see that $\rho(0)$ is sufficiently large when $W=0$.  When disorder increases, the high-energy LLs are broken, leading to the gradually decreasing of $\rho(0)$.  In the sense of renormalization group, sufficiently weak disorder is an irrelevant perturbation in WSM1 \cite{S.Bera} and a marginally relevant perturbation in WSM2.  Here through $\rho(0)$, we further demonstrate that the low-energy LLs are robust to disorder, while the high-energy LLs are easily broken by disorder.  The performance of $\rho(0)$ vs disorder can be used as the third signature to distinguish WSM1 from WSM2.

\section{Conclusions and Discussions}

To summarize, in this paper we have investigated the effect of disorder on the LLs and magnetic transport in tilted WSMs.  By calculating the Hall conductivity and DOS, we find several observable signatures to distinguish WSM1 from WSM2.  The broken of the low-energy LLs in WSM1 by disorder from the higher ones to the lower ones can be understood by the \textit{sink down} picture.  We suggest that our results can be generalized to WSMs with two pairs and even multiple pairs of Weyl nodes. 

Here we have used the quasi-disorder model, while the completely random disorder potential in 3D WSMs may be effectively treated by using the kernel polynomial method \cite{J.H.Wilson, A.Wei}, which needs more work in the future.  There are also some open questions, such as the critical properties of disorder-induced phase transitions in WSM1 under the magnetic field and the effect of disorder on the quantum Hall effect in WSMs, with the physical mechanism originating from the Weyl orbits \cite{T.Schumann, C.Zheng}, or from the topological-insulator-type surface states \cite{B.C.Lin}.  The interplay between disorder, magnetic field and topological Weyl and Dirac semimetals will open up new research into topological phenomena and device applications in three dimensions beyond 2D electron system.

\section{Acknowledgments}

We would like to thank Fuxiang Li, Linghua Wen and W. Vincent Liu for many helpful discussions.  This work was supported by NSFC (Grant No. 11804122) and China Scholarship Council (No. 201706795026).

\section{Appendix} 

\begin{figure}
	\includegraphics[width=8.8cm]{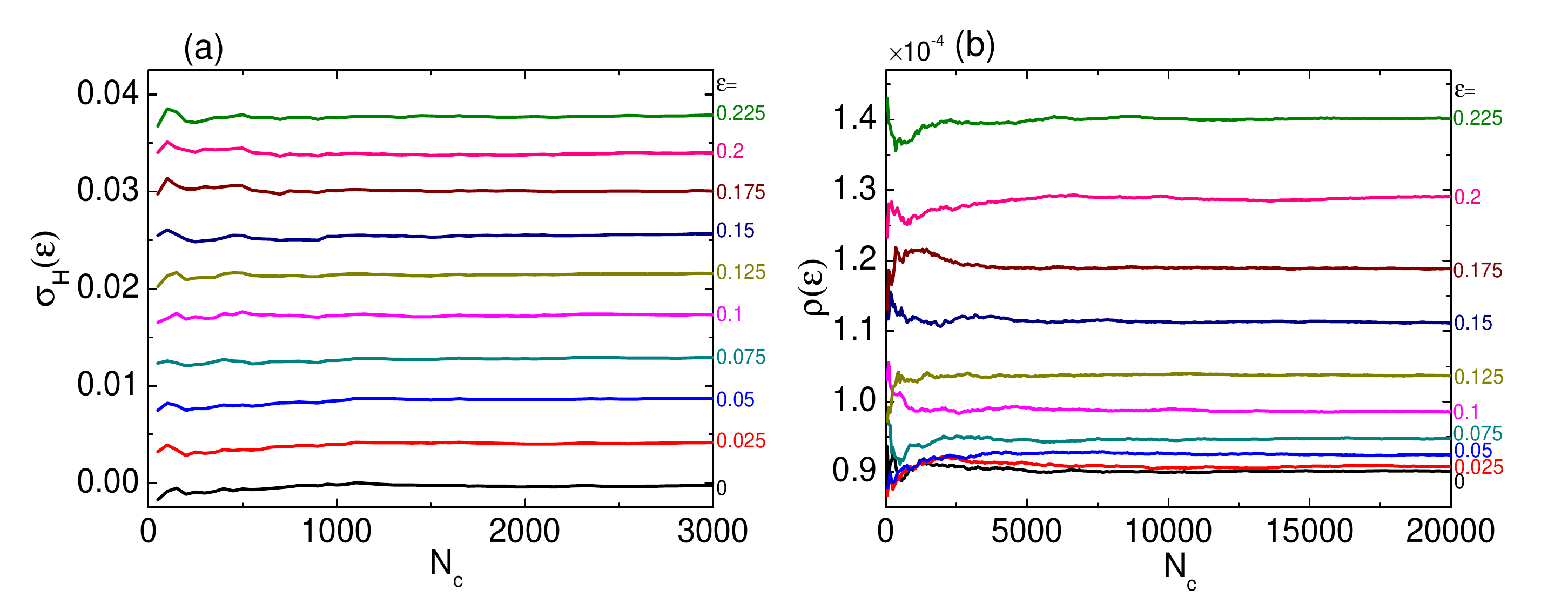}
	\caption{(Color online) (a) The Hall conductivity $\sigma_H(\varepsilon)$ for different energy $\varepsilon$ vs configuration number $N_c$, with $t_z=0$ and $W=4$.  (b) The DOS $\rho(\varepsilon)$ for different energy $\varepsilon$ vs the configuration number $N_c$, with $t_z=0.4$ and $W=3.4$.  We set the magnetic flux $p=60$. }
	\label{Appen}
\end{figure}
 
Here we provide some evidences for the convergence of the results after the configuration number $N_c=10^3\sim10^4$ is performed for the disordered WSM system.  The Hall conductivity $\sigma_H$ and the DOS $\rho$ for different energy $\varepsilon$ vs $N_c$ are shown in figs.~\ref{Appen}(a) and (b), respectively.  We can see that at the beginning, there are certain fluctuations in both physical quantities, but $\sigma_H$ exhibit good convergence when $N_c>2000$ and $\rho$ shows convergence when $N_c>15000$.  This is because $\sigma_H$ is contributed from all the electronic states below the energy $\varepsilon$ while $\rho$ is only due to the electronic states around $\varepsilon$.  As $\sigma_H$ includes much more electronic states than $\rho$, so $\sigma_H$ converges more quickly.  In this paper, we need to take proper configuration number $N_c$ in the calculations as to achieve more precise results of $\sigma_H$ and $\rho$.

\end{document}